\documentclass[journal=jacsat,manuscript=article]{achemso}

\usepackage[version=3]{mhchem} 
\usepackage{upgreek}
\usepackage[nolist]{acronym}
\begin{acronym}
\acro{EDL}{electrical double layer}
\acro{PB}{Poisson--Boltzmann}
\acro{LPB}{linear Poisson--Boltzmann}
\acro{NLPB}{non-linear Poisson--Boltzmann}
\acro{CANDLE}{charge-asymmetric nonlocally determined local electric response}
\acro{AIMD}{\textit{ab initio} molecular dynamics}
\acro{DFT}{density functional theory}
\acro{QM}{quantum mechanics}
\acro{MM}{molecular mechanics}
\acro{MD}{molecular dynamics}
\acro{RISM}{reference interaction site model}
\acro{XRISM}{extended RISM}
\acro{DRISM}{dielectrically-consistent RISM}
\acro{ESM}{effective screening medium}
\acro{LJ}{Lennard-Jones}
\acro{LB}{Lorenz--Berthelot}
\acro{UFF}{universal force field}
\acro{KH}{Kovalenko--Hirata}
\acro{IHP}{inner Helmholtz plane}
\acro{OHP}{outer Helmholtz plane}
\acro{PZC}{potential of zero charge}
\end{acronym}

\newcommand{\qmmm}{QM/MM}
\newcommand{\qmrism}{QM/RISM}
\newcommand{\qmdrism}{QM/DRISM}
\newcommand{\qmesmrism}{QM/ESM-RISM}
\newcommand{\esmrism}{ESM-RISM}

\newcommand{\vaspsolpp}{VASPsol++}

\SectionNumbersOn

\author{Alessandro Mangiameli}
\affiliation[TUM]{Department of Chemistry and Catalysis Research Center, TUM School of Natural Sciences, Technical University Munich, 85748 Garching, Germany}
\author{Christopher J. Stein}
\affiliation[TUM]{Department of Chemistry and Catalysis Research Center, TUM School of Natural Sciences, Technical University Munich, 85748 Garching, Germany}
\alsoaffiliation[AMC]{Atomistic Modeling Center, Munich Data Science Institute, 85748 Garching, Germany}
\email{christopher.stein@tum.de}

\title[An \textsf{achemso} demo]
  {Investigating the Electrochemical Double Layer with Quantum-Chemical Simulations and Implicit Solvation Models}

\abbreviations{IR,NMR,UV}

\begin{document}

\begin{tocentry}

\includegraphics[width=8.5cm]{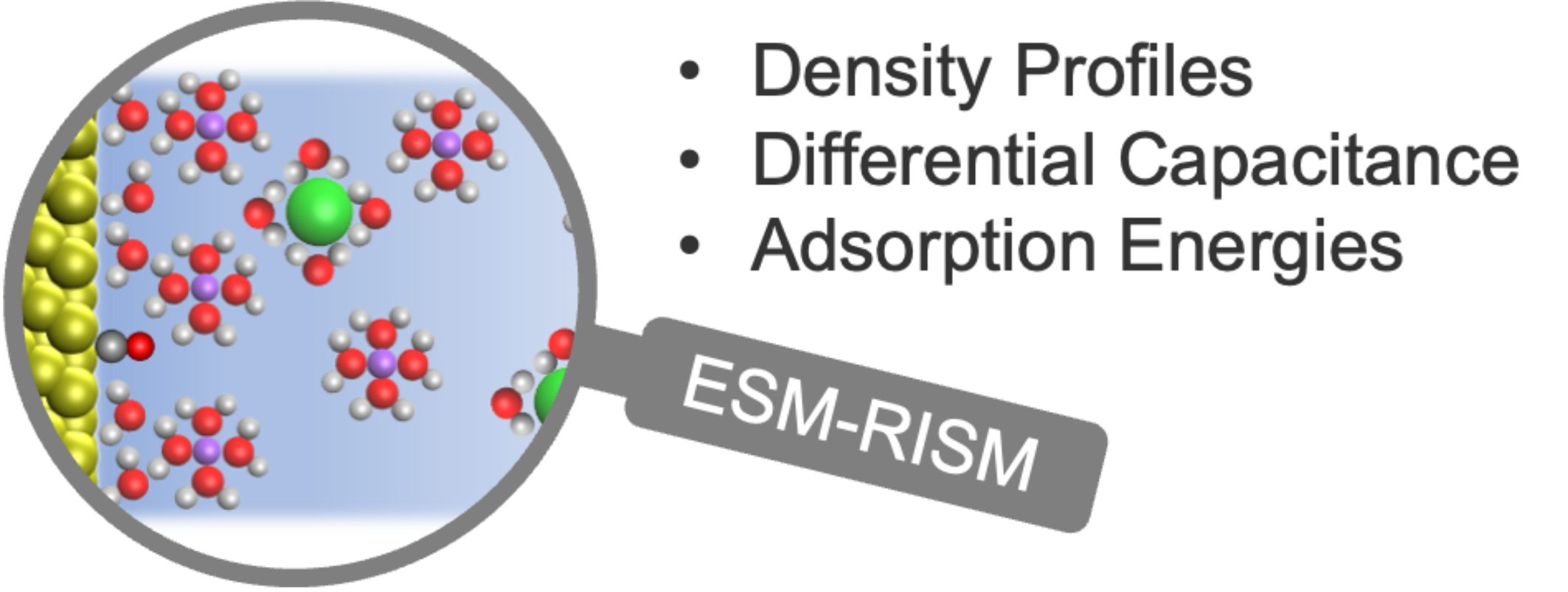}






\end{tocentry}

\begin{abstract}

We assess the dielectrically consistent reference interaction site model (DRISM) as an implicit electrolyte framework for modeling the electrochemical double layer, and compare it with the Poisson--Boltzmann model and explicit molecular dynamics results from the literature.
We use the gold--electrolyte interface as the main test case and analyze solvent and ionic density profiles, the differential capacitance, and the solvation contribution to $\ce{CO}$ adsorption.
The results show a strong sensitivity to the Lennard-Jones parametrization of metal--ion and metal--water interactions.
In particular, we find that the default Lorentz--Berthelot mixing rules to be inadequate and lead to excessive $\ce{Na+}$ accumulation at the interface, which results in an increase of the differential capacitance at negative electrode potentials.
We demonstrate that introducing pair-specific metal--ion parameters yields more symmetric charging behavior and provides greater flexibility.
Our findings suggest that using pair-specific parameters, rather than relying on Lorentz--Berthelot mixing rules, improves the accuracy of the model and opens the way for future studies with this improved yet equally performant model.

\end{abstract}

\section{Introduction}
\label{introduction}

The structure and properties of the \ac{EDL} at charged interfaces play a central role in electrochemistry.
From energy storage in supercapacitors to electrocatalysis and corrosion,
the interfacial region controls ion distributions,
potential profiles, and ultimately the rates and selectivity of electrochemical processes \cite{dengInterfacialElectrolyteEffects2022,schottHowAssessPredict2024, shinImportanceElectricDouble2022a}.
A quantitative understanding of the \ac{EDL} is therefore essential for the design of efficient electrochemical systems.

Atomistic simulations of interfacial phenomena must adequately model the intricate interplay among solvent,
ions, and charged surfaces.
The most common theoretical framework for \ac{EDL} is the \ac{PB} model originally
developed by Gouy \cite{gouyConstitutionChargeElectrique1910a} and
Chapman \cite{chapmanLIContributionTheory1913a} and
later by Stern \cite{sternZURTHEORIEELEKTROLYTISCHEN1924b}.
Its mean-field formulation, in which ions are treated as point charges in a continuum dielectric,
allows for simple, computationally inexpensive evaluation of ion density profiles and
potential distributions.
Thanks to its simplicity, the PB model (and related models \cite{ringeImplicitSolvationMethods2022}) is,
even today, one of the most common approaches, and it is implemented in various well-known
electronic structure software packages
\cite{mathewImplicitSolvationModel2014, sundararamanChargeasymmetricNonlocallyDetermined2015, mathewImplicitSelfconsistentElectrolyte2019, nattinoContinuumModelsElectrochemical2019, steinPoissonBoltzmannModel2019}.
Nevertheless, its approximations neglect molecular granularity and correlation effects,
which become critical at high ionic strength or in the immediate vicinity of electrodes.

A much more detailed picture can be obtained from first-principles methods like \ac{AIMD},
where the electrolyte is modeled explicitly at the \ac{DFT} level
\cite{heenenSolvationMetalWater2020, goldsmithEffectsAppliedVoltage2021, xuBenchmarkingWaterAdsorption2024a, raffoneRevealingMolecularInterplay2025}.
\ac{AIMD} is, in principle, providing the most accurate description;
however, it is computationally very demanding and therefore limits the size of the simulation box and
the time scale of the simulation.
To reduce the computational cost, the electrolyte is often treated classically in a \ac{QM}/\ac{MM} framework
\cite{shinImportanceElectricDouble2022a}.
But even with this substantially decreased computational cost, sufficient sampling is challenging
since long equilibration times and large simulation cells are usually required \cite{schwarzElectrochemicalInterfaceFirstprinciples2020}.
This becomes especially challenging in constant-potential simulations,
where specialized algorithms to control the electrode potential while maintaining charge neutrality need to be applied \cite{schwarzElectrochemicalInterfaceFirstprinciples2020}.

An alternative approach is provided by treating the electrolyte with the \ac{RISM}.
\ac{RISM}, first introduced by \citet{chandlerOptimizedClusterExpansions1972}, is an approximation to solve the molecular Ornstein--Zernike equation and obtain the site-site radial distributions functions for the given system;
it has been extended to molecular fluids with charged sites by \citet{hirataExtendedRismEquation1981}
also indicated as \ac{XRISM} and then further developed for electrolytes
by Perkyns \cite{perkynsDielectricallyConsistentInteraction1992, perkynsSiteSiteTheory1992},
who introduced a dielectrically consistent variant which apport a correction to accounts for the solvent polarizability.
\ac{RISM} has been first embedded to \ac{QM} calculations by \citet{satoSelfconsistentFieldInitio2000} and further extended for electrochemical simulations
by \citet{nishiharaHybridSolvationModels2017} and combined with the \ac{ESM},
\cite{otaniFirstprinciplesCalculationsCharged2006} in short \esmrism,
which effectively removes the periodicity in the direction perpendicular to the slab,
enabling a non-repeated vacuum/slab/solvent configuration.
The \qmesmrism{} framework has been employed in various studies
\cite{haruyamaElectrodePotentialDensity2018, teschPropertiesPt111Electrolyte2021a,weitznerEngineeringSolutionMicroenvironments2020,sudhakarInterfacialElectricFields2024}
and proved to yield essential insights into the structure of the electrode--electrolyte interface.
In two recent works, \esmrism{} has been compared to the linearized \ac{PB} model
for studying of the $\ce{CO}$ reduction reaction at the $\ce{Cu}(100)$ facet
\cite{pasumarthiComparativeStudyElectrical2023}
and with the \ac{CANDLE} model \cite{sundararamanChargeasymmetricNonlocallyDetermined2015}
(which also belongs to the family of the linearized PM models)
for the studying of the standard hydrogen electrode \cite{demeyereComparisonModernSolvation2024}.
Both studies report that \esmrism{} generally leads to destabilization of
the adsorption energies within the employed parametrizations.
In the second study, \citeauthor{demeyereComparisonModernSolvation2024} compared the differential capacitance
computed with \ac{CANDLE} and \esmrism{}, noticing considerable differences between the two models. 

In this work, we extend the previous studies by further evaluating the capabilities of \qmesmrism{}
and of the newly developed dielectrically consistent version \ac{DRISM}
\cite{hagiwaraDevelopmentDielectricallyConsistent2022},
for the modeling of the electrode-electrolyte interface.
In the first part, we focus on the solvent and ionic density profiles at an $\ce{Au}(111)$ facet.
In the second part, we study the differential capacitance and its dependence on the
electrolyte parametrization, compositions, and concentrations,
since this quantity can be compared rather directly to experimental studies.
Finally, we test the impact of the parametrization for the adsorption energy of $\ce{CO}$ at the $\ce{Au}$ surface.
We systematically compare the \ac{DRISM} results with the recently implemented
non-linear \ac{PB} model \vaspsolpp{}. \cite{islamImplicitElectrolyteModel2023}

\section{Results and Discussion}
\label{results_discussion}

\subsection{Density Profiles}
\label{Density Profiles}

The most direct outputs of a \qmdrism{} calculation are the classical particle density distributions
for the implicit \ac{DRISM} electrolyte around the explicit QM system.
The solvent and ionic density profiles are the $x$- and $y$-averaged particle densities
as a function of the slab distance $z$.
Since these profiles cannot be directly measured experimentally,
computational models such as \ac{DRISM} are valuable tools for elucidating the structure of the \ac{EDL}.

\subsubsection{Solvent density profiles}
\label{Solvent density profile}

The structure of the metal--water interface has been extensively investigated. 
For non-absorbing metals like $\ce{Au}$, water is known to form a layer
in which the water molecules lie parallel to the surface
\cite{nadlerEffectDispersionCorrection2012,heenenSolvationMetalWater2020,alfaranoStrippingAwayIon2021}.
For adsorbing metals like $\ce{Pt}$, the peak for the oxygen appears split,
due to its specific adsorption on the surface \cite{heenenSolvationMetalWater2020}.
As already reported by \citet{demeyereComparisonModernSolvation2024},
\qmrism{} is unable to predict the double-peak structure,
given the lack of specific chemical bonding interactions in the metal--water potential,
and we would expect the same deficiency in a fully classical \ac{MD} or a \qmmm{} simulation,
where the electrolyte is modeled classically.

\begin{figure}[!htb]
\begin{center}
\includegraphics[width=8.6cm]{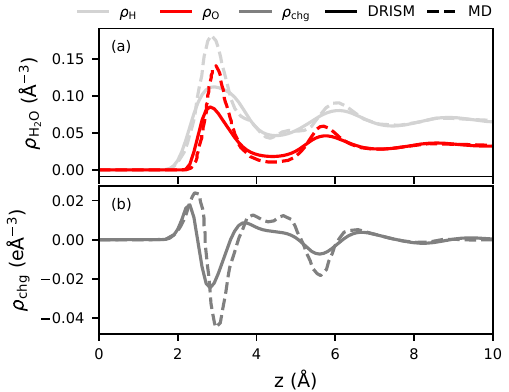}
\caption{
Panel (a) displays hydrogen (light gray) and oxygen (red) particle density profiles
computed with \ac{DRISM} (continuous line) and explicit \ac{MD} simulations (segmented line)
[taken from Hughes et al.\citenum{hughesStructureElectricalDouble2014}] 
for the $\ce{Au}(111)$ facet in water.
The mTIP3P water model is used in both cases.
For the \ac{DRISM} calculation, we adjusted the LJ parameter for $\ce{Au}$
to match the $\ce{Au}$-$\ce{O}$ GolP parametrization. 
Panel (b) shows the respective charge density profiles.
}
\label{DRISM_vs_MD_Au_111_H2O}
\end{center}
\end{figure}

To validate the water density profiles predicted by \ac{DRISM} for the less reactive $\ce{Au}(111)$ facet,
we compare them with the particle and charge profiles obtained from
the classical \ac{MD} simulations of \citet{hughesStructureElectricalDouble2014} (Fig.~\ref{DRISM_vs_MD_Au_111_H2O}).

The mTIP3P water model is used in both cases.
We highlight, however, that an exact correspondence is not expected,
since the classical \ac{MD} \cite{hughesStructureElectricalDouble2014} is performed with the GolP-CHARMM force field,
which uses dipole sites on the gold surface to simulate the metal polarization,
while in the \qmdrism{} embedding, the polarization is taken into account directly by the density functional.
For the hydrogen and oxygen density profiles, the positions of the first and second peaks are in good agreement;
however, the height of the peaks predicted by \ac{DRISM} is lower;
this tendency might also be ascribed to the \ac{KH} closure,
which has been reported to underestimate the height of the correlation peaks \cite{perkynsProteinSolvationTheory2010}.
As a consequence, the peaks in the charge density profiles are also lower in \ac{DRISM} than in \ac{MD}.
Despite these differences, we observe a generally good agreement between \ac{DRISM} and the classical \ac{MD} simulations.

For completeness, we also mention that, while for \ac{DRISM} and classical \ac{MD}s,
there is a one-to-one mapping between the particle density profiles and the charge density profiles,
this is not the case for \ac{AIMD}, where the electronic charge is obtained from \ac{DFT};
therefore, even for similar particle density profiles, the charge density profiles might be slightly different.

When a potential is applied to the electrode, water molecules are tilted by electrostatic interactions,
resulting in a shift in the hydrogen and oxygen peaks.
This behavior is observed in both classical \cite{parkHelmholtzCapacitanceAqueous2022}
and \textit{ab initio} calculations \cite{goldsmithEffectsAppliedVoltage2021}
as well as in \qmmm{} \cite{shinImportanceElectricDouble2022a},
and has already been documented for \qmrism{} \cite{demeyereComparisonModernSolvation2024}.
However, a quantitative comparison between the \qmrism{} and \ac{MD} density profiles at finite applied potential
is more complicated due to two reasons: first the different definitions of the electrode potential in the two methods,
which is based on the Fermi level referenced to the electrolyte bulk potential in \qmrism{} and on the surface charge in classical \ac{MD}, respectively,
and second the important role of their distinct treatments of metal polarization.

\begin{figure}[!htb]
\begin{center}
\includegraphics[width=8.6cm]{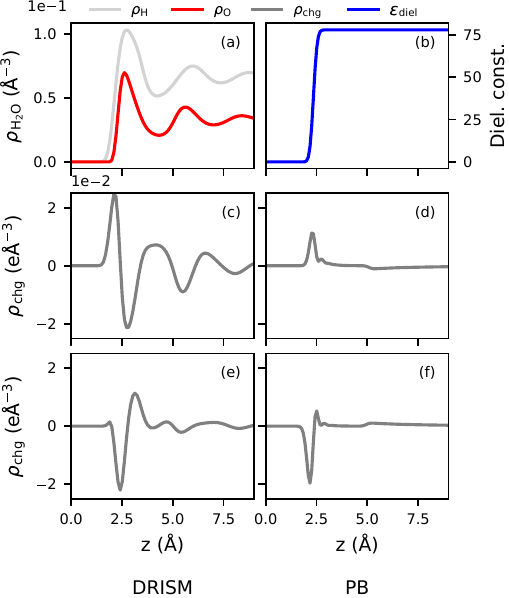}
\caption{
Comparison between solvent modeling in \ac{PB} and \ac{DRISM}.
In (a), the particle density profiles of water computed by \ac{DRISM} are shown,
whereas in the \ac{PB} model, the solvent is replaced with a continuum dielectric.
In (c) and (d) are shown the charge density profile from \ac{DRISM}
and the bound charge from the \ac{PB} for
a negative electrode surface charge $\sigma=-8.3\,\upmu\text{C/cm}^{-2}$ potential with respect to the \ac{PZC}.
In (e) and (f), the same applies for a positive electrode potential.
}
\label{solvent_charge_density_DRISM_vs_PB}
\end{center}
\end{figure}

We now turn to the comparison between \ac{DRISM} and the \ac{PB} model. 
In Fig.~\ref{solvent_charge_density_DRISM_vs_PB},
we show a side-by-side comparison of the two models for both negative $\sigma=-8.3\,\upmu\text{C}\text{/cm}^{-2}$
and positive $\sigma=+8.3\,\upmu\text{C}\text{/cm}^{-2}$ surface electrode charge,
which correspond to a negative and positive applied potential relative to the \ac{PZC}.

In the \ac{PB} model, the solvent is modeled as a dielectric continuum, as shown in panel (b).
In the presence of the solute, the dielectric medium generates a polarization field $\mathbf{P}(\mathbf{r})$;
the bound (or polarization) charge is then defined as
$\rho_{\mathrm{b}}(\mathbf{r}) = - \nabla \mathbf{P}(\mathbf{r})$.
As we can see, while the charge profile predicted by \ac{DRISM} exhibits complex oscillatory behavior,
the bound charge obtained from the \ac{PB} model shows mainly one peak
situated at the boundary of the dielectric cavity,
followed by an exponential peak due to explicit ionic charge screening.

%
%
%
%
%
%
%
%
%
%
%
%
\subsubsection{Ionic density profiles}
\label{Ionic density profiles}

\begin{figure*}[!htb]
\begin{center}
\includegraphics[width=14cm]{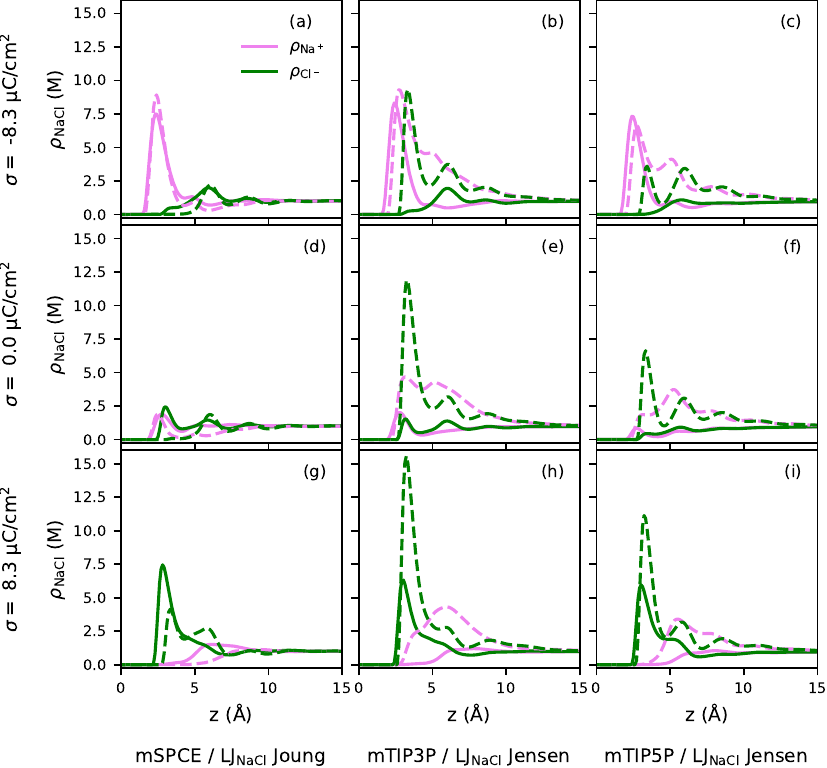}
\caption{
Ionic density profiles for the $\ce{Au}(111)$ facet in contact with $1\,\text{M}$ $\ce{NaCl}$.
(a), (b) and (c) negative surface charge $\sigma = - 8.3\,\upmu\text{C/cm}^{2}$;
(d), (e) and (f) neutral electrode $\sigma = 0\, \upmu\text{C/cm}^{2}$;
(g), (h) and (i) $\sigma = + 8.3\,\upmu\text{C/cm}^{2}$.
The continuous lines refer to the UFF parametrization for the $\ce{Au}$ electrode,
while the segmented line refers to the parametrization from \citet{heinzAccurateSimulationSurfaces2008}.
}
\label{DRISM_Ionic_density_profiles_1.00M}
\end{center}
\end{figure*}

We now move our attention to the ionic density profiles.
A central step in the \qmrism{} framework is selecting the pair-potential parameters.
In the literature, numerous \ac{LJ} parametrizations for water--ion systems have been proposed,
each optimized for different thermodynamic or structural properties
\cite{joungDeterminationAlkaliHalide2008, jensenHalideAmmoniumAlkali2006, blazquezMadrid2019ForceField2022}.
We discuss the $\ce{Au}(111)$ facet as a common catalytic surface in contact with aqueous $\ce{NaCl}$ as a typically used electrolyte solvent.
In Fig.~\ref{DRISM_Ionic_density_profiles_1.00M} we present the ionic density profiles for an electrolyte concentration of $1\,\text{M}$,
obtained (from top to bottom) at negative, neutral, and positive surface charge, respectively.

For the electrolyte, we use three different water/ions parametrizations:
the mSPCE water model with the ionic parametrization from \citet{joungDeterminationAlkaliHalide2008} (left),
the mTIP3P water model with the ionic parametrization from \citet{jensenHalideAmmoniumAlkali2006} (middle),
and the mTIP5P water model, again with the parametrization from \citet{jensenHalideAmmoniumAlkali2006} (right).
For the gold surface, we adopt the \ac{LJ} parametrization from \citet{heinzAccurateSimulationSurfaces2008} (segmented lines),
which was specifically optimized for classical simulations of the metal--water interface,
and the \ac{UFF} from \citet{rappeUFFFullPeriodic1992} (continuous line).

Apparently, the \ac{UFF} parametrization for the gold surface
yields relatively low variability in the density profiles for the three water models. 
When we adopt the Heinz parametrization, the profiles change significantly,
which reflects a considerably stronger metal--electrolyte interaction, with the \ac{LJ} well depth parameter $\varepsilon_{\mathrm{Au}}$ = $5.29\,\text{kcal/mol}$
compared to $\varepsilon_{\mathrm{Au}}$ = $0.039\,\text{kcal/mol}$ in UFF.
We recall that under the default (\ac{LB}) mixing rules, the cross terms are generated as the geometric mean $\varepsilon_{ij} = \sqrt{\varepsilon_i \varepsilon_j}$, therefore, changing the gold self-parameters directly propagates to all $\ce{Au}$--water and $\ce{Au}$--ion interactions.

Additionally, we observe that the cation peak is located between $2$ and $3\,\text{\AA}$ from the metal surface,
i.e., within the same region as the oxygen peak of the first water layer.
This indicates that the cations are found directly at the \ac{IHP}.
The cation accumulation becomes more pronounced at negative electrode charge
and we attribute this as the cause of the high differential capacitance, as discussed in Sec.~\ref{differential capacitance}.

%
%
%
%
%
%
%
%
%
%
%
%
%
%
%
%
%
This tendency for $\ce{Na+}$ to accumulate directly at the surface is not specific to \ac{DRISM} but is also observed in explicit molecular dynamics simulations.
In a recent experimental and theoretical work from \citet{alfaranoStrippingAwayIon2021},
it was found that $\ce{Na+}$ and $\ce{Cl-}$ behave asymmetrically at the $\ce{Au}$-water interface.
Under negative bias, $\ce{Na+}$ enters the \ac{IHP} already at very small potentials and partially loses one hydration water,
not because of strong $\ce{Na+}$-$\ce{Au}$ bonding,
but because the applied negative field disrupts the interfacial 2D hydrogen-bond network,
making insertion energetically favorable.
In contrast, under positive bias, $\ce{Cl-}$ remains fully hydrated in the \ac{OHP} over a broad potential range and only desolvates at high positive potentials.

\begin{figure}[!htb]
\begin{center}
\includegraphics[width=8.6cm]{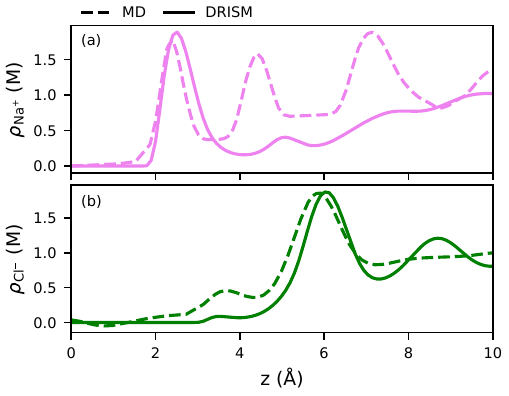}
\caption{
Density profiles for the $\ce{Au}(111)$ facet in contact with $1\,\text{M}$ $\ce{NaCl}$ in water.
The profiles represented as dotted lines are taken from a classical \ac{MD} simulation of \citet{ntimMolecularDynamicsSimulations2023} using a polarizable force field for $\ce{Au}$.
The continuous profiles are obtained from \ac{DRISM} using the mSPCE water model,
the Heinz parametrization for the $\ce{Au}$ slab, and the \citeauthor{joungDeterminationAlkaliHalide2008} parametrization for the ions.
In both cases, the electrode is at the \ac{PZC}.
Although the parametrization of the pair potentials is not identical,
there is qualitative agreement in the positions of the first peaks for sodium and chloride.
}
\label{DRISM_vs_MD_Au_111_NaCl_1M}
\end{center}
\end{figure}

Given the approximations on which \ac{DRISM} is based, we cannot expect it to capture the exact interfacial 2D hydrogen-bond network and its disruption under applied potential.
However, we still consider the ionic density profiles obtained from explicit \ac{MD}s as a useful benchmark for the model.
In Fig.~\ref{DRISM_vs_MD_Au_111_NaCl_1M},
we compare the density profiles obtained from a classical \ac{MD} simulation of the $\ce{Au}(111)$ interface in contact with $1\,\text{M}$ $\ce{NaCl}$ at the \ac{PZC} by \citet{ntimMolecularDynamicsSimulations2023} with the corresponding \ac{DRISM} results.

As shown in panel (a), the position of the first $\ce{Na+}$ peak agrees well between the \ac{DRISM} and \ac{MD} results.
Specific cation adsorption has also been observed by \citet{servaEffectMetallicityCapacitance2021} for the $\ce{Au}(100)$ facet, where the adsorption strength was found to increase with the metallic character of the electrode.
Interestingly, for the anion (panel (b)), we observe the main density peak at $\approx 6\,\text{\AA}$, in correspondence with the second water layer, and a minor peak at $\approx 3.5\,\text{\AA}$.
The positions of both peaks are in qualitative agreement with the \ac{MD} simulation.
Also in this case, there is an agreement with a similar profile by \citet{servaEffectMetallicityCapacitance2021}.

From the classical theory of the electrical double-layer \cite{grahameElectricalDoubleLayer1947}, however,
small cations such as $\ce{Li+}$ and $\ce{Na+}$ are known to typically retain their solvation shell and reside at the \ac{OHP}, rather than populating the \ac{IHP}, directly at the surface.

One potential limitation of the \qmrism{} implementation in Quantum Espresso is the use of \ac{LB} mixing rules to construct the cross pair-potential parameters \cite{nishiharaHybridSolvationModels2017}. 
In the GolP-CHARMM force field \cite{wrightGolPCHARMMFirstPrinciplesBased2013},
which is specifically parameterized for the $\ce{Au}$-water interface,
the mixing rules are applied only to generic species,
whereas the metal-water interactions are treated with pair-specific parameters fitted to reproduce \ac{DFT} adsorption energies of a single water molecule.
In the classical \ac{MD} studies of \citet{hughesStructureElectricalDouble2014} and \citet{parkHelmholtzCapacitanceAqueous2022}, both employing GolP-CHARMM, no cation accumulation directly at the interface is observed.
In another recent work from \citet{shinImportanceElectricDouble2022a}, for an $\ce{Ag}(111)$ electrode, $\ce{Na+}$ does not exhibit specific adsorption at cathodic potentials but remains solvated in the \ac{OHP}, preserving its first hydration shell (coordination number  $\approx$ 6) and avoiding direct contact with the $\ce{Ag}(111)$ surface. 
In this case, the metal-electrolyte interaction was modeled using first-principles-derived Buckingham-type van der Waals potentials for $\ce{Ag}$-water and $\ce{Ag}$-ion interactions.
We also note that an extensive study by \citet{bergEvaluationOptimizationInterface2017} concluded that the \ac{LJ} pair potential itself is insufficient for the $\ce{Au}$-water interaction, and that alternative potentials such as the Morse or Buckingham potentials are more suitable options.
Considering the different observations and conclusions of the studies discussed above,
the microscopic structure of the electrical double layer at metal-aqueous interfaces seems still not fully understood, and several details remain open for debate.

It is therefore not straightforward to establish a unique benchmark for the \ac{DRISM} results.
For this reason, we proceed by examining the \ac{EDL} structure predicted by \ac{DRISM},
when pair-specific parameters are used, instead of the default \ac{LB} mixing rules.

\begin{figure}[!htb]
\begin{center}
\includegraphics[width=8.6cm]{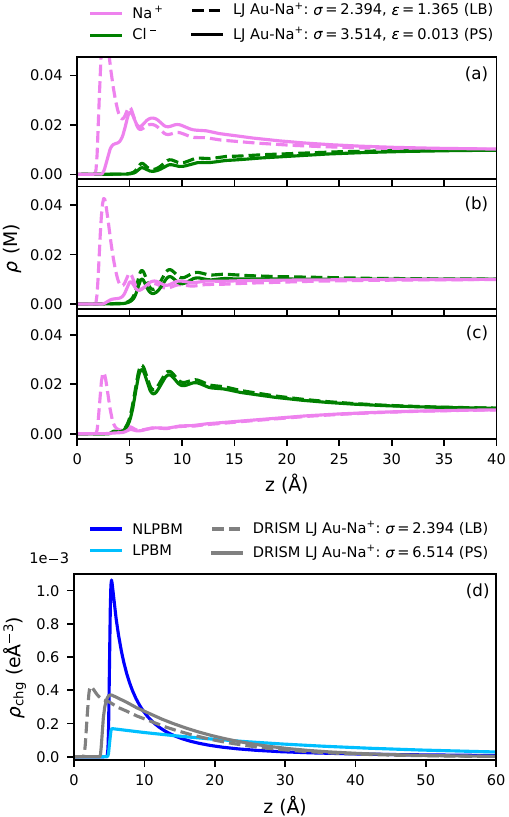}
\caption{
Density profiles for $\ce{Au}(111)$ facet in contact with $0.01\,\text{M}$ electrolyte.
The calculations are performed at a surface charge of $\sigma = -0.27\,\upmu \text{C}/\text{cm}^{2}$, \ac{PZC}, $\sigma = +0.27\,\upmu \text{C}/\text{cm}^{2}$, and $\sigma = -8.33\,\upmu \text{C}/\text{cm}^{2}$ in (a), (b), (c), and (d) respectively.
The segmented lines are computed with the $\ce{Au}$ Heinz / mSPCE / $\ce{NaCl}$ Joung parameterization and the default \ac{LB} mixing rules.
For the continuous lines, pair-specific parameters are given for the $\ce{Au}$-$\ce{Na}$ \ac{LJ} interaction.
In (d) results from the non-linear and linear \ac{PB} model are also shown for comparison.
}
\label{Stern_and_diffuse_layer}
\end{center}
\end{figure}

%
%
%
%
%
%
%
%
%
%
%
%
In Fig.~\ref{Stern_and_diffuse_layer}~(a) we compare the density profiles at the \ac{PZC} and for a small electrode surface charge obtained at a lower electrolyte concentration of $0.01\,\text{M}$ with the $\ce{Au}$ Heinz / mSPCE $\ce{NaCl}$ Joung parametrization.
For the segmented curves, the default \ac{LB} mixing rules are used and the sodium ion exhibits a defined peak within the \ac{IHP}.
This peak is also present for a positive surface charge applied of $\sigma = + 0.27\,\upmu \text{C}/\text{cm}^{2}$, in panel (c);
at positive electrode charge we expect the electrostatic repulsion to dominate and we therefore consider this peak as unphysical.
By selectively decreasing the \ac{LJ} interaction (see Fig.~S2) between gold and sodium (continuous line), we observe, as expected, the disappearance of the first sodium peak.

These results suggest that \ac{DRISM} is, in principle, capable of predicting the exclusion layer at the interface.
As discussed before, the formation of such a layer arises from a complex interplay between the surface electronic structure,
specific ion-surface interactions, and the rearrangement of the hydrogen-bond network in the interfacial solvent.
Within the \qmdrism{} framework, these effects are incorporated only at the level of site-site correlation functions and approximate closure relations.
Therefore, the emergence of an exclusion layer may reflect an effective description rather than a fully faithful representation of the underlying molecular mechanisms.
Nevertheless, from a modeling perspective, the parameters could be empirically tuned to reproduce desired interfacial properties, such as ion depletion or trends in capacitance.

Within the \ac{PB} family of models, the exclusion layer is typically introduced through an empirical ionic cavity correction (also referred to as a Stern-layer correction)
\cite{steinPoissonBoltzmannModel2019,nattinoContinuumModelsElectrochemical2019,islamImplicitElectrolyteModel2023,sundararamanImprovingAccuracyElectrochemical2018}.
A similar strategy can be used in \ac{DRISM} by increasing the effective \ac{LJ} repulsion radius between the electrode and the ions, which shifts ions away from the surface and generates an exclusion layer.
Figure~\ref{Stern_and_diffuse_layer}~(d) compares this behavior with the linear and non-linear \ac{PB} models.
Increasing the $\ce{Au}$-$\ce{Na+}$ \ac{LJ} repulsion radius indeed produces an exclusion layer analogous to the Stern-layer correction in \ac{PB}.

Although \ac{PB} cannot reproduce the molecular-scale structure of the compact layer, it is expected to describe the diffuse region more reliably in the dilute limit.
In this regime, the electrolyte approaches an ideal gas of point ions in a dielectric continuum, and the ionic density decays exponentially away from the surface.
As we can see \ac{DRISM} reproduce the expected exponential decay at large distances from the surface,
with a charge density which is intermediate between the linear and non-linear \ac{PB} models.

%
%
%
%
%
%
%
%
%
%
%
%
%
%
%
\subsection{Differential Capacitance}
\label{differential capacitance}

The differential capacitance quantifies how the surface charge responds to changes in the applied potential.
It has been extensively studied and remains one of the primary experimental probes of the structure of the electrical double layer.

\begin{figure}[htb]
\begin{center}
\includegraphics[width=8.6cm]{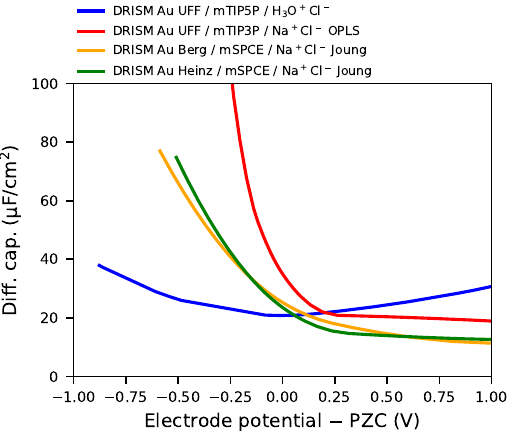}
\caption{
Differential capacitance for the $\ce{Au}(111)$ facet in contact with $1\,\text{M}$ aqueous electrolyte. For $\ce{H3O+}$ (blue curve), the differential capacitance behaves as previously reported \cite{demeyereComparisonModernSolvation2024}.
When $\ce{Na+}$ is used as a cation, the strong peak discussed in the previous section causes a divergence in the computed differential capacitance at negative electrode potentials.
}
\label{diff_cap_DRISM_Au_1.00M}
\end{center}
\end{figure}

In Fig.~\ref{diff_cap_DRISM_Au_1.00M}, we report the differential capacitance as a function of the electrode potential relative to the \ac{PZC} for the $\ce{Au}(111)$ surface in $\ce{NaCl}$ and $\ce{HCl}$ ($\ce{H3O+Cl-}$) electrolytes at $1\,\text{M}$, using different parameters.
For $\ce{NaCl}$, we find that the differential capacitance increases steeply at negative electrode potentials.

Experimentally, this steep increase in the differential capacitance is observed at positive electrode potentials and is ascribed to specifically adsorbed anions \cite{adnanTrackingSurfaceStructure2024a, valetteDoubleLayerSilver1989},
however the measurements are typically done between $0.01\,\text{M}$ and $0.1\,\text{M}$ concentration.
In the study of \citet{devanathanDifferentialCapacitanceSolid1973},
differential capacitance was measured at $1\,\text{M}$ concentrations for different metals and different electrolytes;
capacitance spikes at negative electrode potentials were also not observed.
Furthermore, for a non-adsorbing electrolyte, the differential capacitance is expected to exhibit a hump near the \ac{PZC} at high electrolyte concentrations \cite{sundararamanImprovingAccuracyElectrochemical2018}.

We therefore regard this excessively large differential as unphysical,
and we trace it back to the strong peak in the sodium density profile discussed in the previous section \ref{Ionic density profiles}.
Notably, the $\ce{Au}$ UFF / mTIP3P / $\ce{NaCl}$ Jensen parametrization, which corresponds to the weakest $\ce{Au}$-$\ce{Na+}$ \ac{LJ} interaction ($\varepsilon = 0.0044,\text{kcal/mol}$), yields the largest capacitance.
This result highlights the important role of the $\ce{Au}$-$\ce{H2O}$ interaction in preventing excessive accumulation of $\ce{Na+}$ at the interface.
In the $\ce{Au}$ UFF / mTIP3P parametrization the $\ce{Au}$-$\ce{H2O}$ interaction is relatively weak, whereas it is stronger in the $\ce{Au}$ Heinz / mSPCE parametrization.
We also tested the pair-specific $\ce{Au}$-$\ce{H2O}$ \ac{LJ} parametrization optimized by \citet{bergEvaluationOptimizationInterface2017}.
Although this model employs $\ce{Au}$-$\ce{O}$ \ac{LJ} parameters similar to those of the Heinz parametrization,
it introduces a predominantly repulsive $\ce{Au}$-$\ce{H}$ interaction.
Nevertheless, the resulting differential capacitance is largely unchanged.
The Heinz and Berg parametrizations lead, however, to different solvation contributions in the adsorption energy of $\ce{CO}$,
as discussed in section \ref{Adsorption Energies}.

We mention that the increase of the capacitance at negative potentials was observed in a previous study employing the \qmesmrism{} framework, by \citet{karnesHybridQuantumClassical2022} for the $\ce{Cu}(111)$ and $\ce{Cu}(100)$ facets in contact with the mTIP5P water and different electrolyte ion species,
where they show a clear asymmetry in the charging curves with respect to the \ac{PZC}
(see their supplementary information, and Fig.~S4 where we replot the charging curve and the respective differential capacitance).
In Fig.~S3, we additionally show the differential capacitance obtained using the \ac{LJ} parameters of \citet{haruyamaElectrodePotentialDensity2018} for $\ce{Pt}$,
which were fitted to reproduce the interaction energy between a $\ce{Pt}$ surface and a $\ce{Xe}$ atom from a \ac{DFT} calculation.
Even with these parameters, we still observe pronounced cation accumulation at the interface,
leading to divergent differential capacitance values at negative electrode potentials.

In contrast, when $\ce{H3O+}$ is used as the cation in the \ac{DRISM} electrolyte, this behavior is not observed.
In \ac{DRISM}, $\ce{Na+}$ is represented as a single-site classical particle with its positive charge located at the ion center, which allows it to approach the electrode surface closely.
$\ce{H3O+}$, instead, is modeled as a multi-site species with the positive charge distributed over the three hydrogen atoms.
As a result, the effective charge cannot approach the interface as closely, leading to a smaller accumulation of positive charge at the electrode surface.
In agreement with \citet{demeyereComparisonModernSolvation2024}, we also observe fluctuations in the computed Fermi level,
particularly for simulations using the mTIP5P water model and the $\ce{H3O+}$ cation
and we found the convergence significantly more difficult,
requiring and a convergence threshold for the 3D-RISM solver of $10^{-6}\,\text{Ry}$ instead of $10^{-7}\,\text{Ry}$, as reported in the methods section \ref{methods}.
Moreover, we report that when using the mTIP5P model and the $\ce{H3O+}$ cation with \ac{DRISM}, some calculations did not converge.
This behavior may be linked to the larger number of interaction sites in both mTIP5P water and $\ce{H3O+}$, which increases the complexity of the site-site correlations.

\begin{figure}[htb]
\begin{center}
\includegraphics[width=8.6cm]{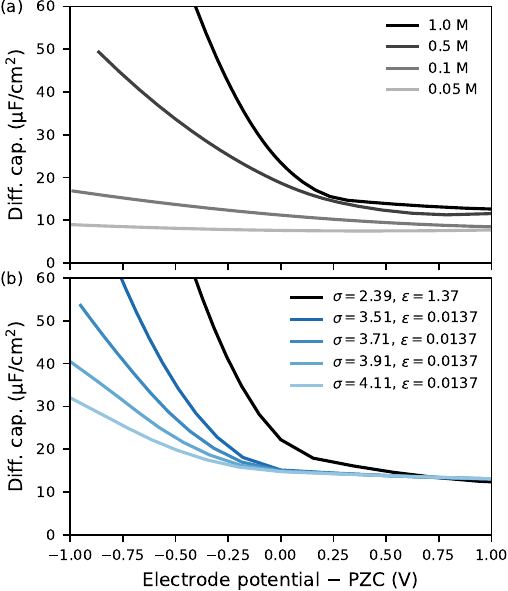}
\caption{
Differential capacitance for the $\ce{Au}(111)$ facet in contact with aqueous $\ce{NaCl}$ electrolyte:
(a) at different electrolyte concentrations;
(b) with different pair-specific parameters for the $\ce{Au}$-$\ce{Na+}$ interaction.
The black curve corresponds to the default \ac{LB} mixing rules, while for the other curves,
the $\ce{Au}$-$\ce{Na+}$ parameters are changed as in the legend.
The remaining parameters are obtained using the default mixing rules for the $\ce{Au}$ Heinz / mSPCE / $\ce{NaCl}$ Joung parametrization.
}
\label{diff_cap_DRISM_PS_and_conc_Au_NaCl}
\end{center}
\end{figure}

%
%
%
%
%
%
%
%
%
%
%
%
%
%
%
%
%
%
In Fig.~\ref{diff_cap_DRISM_PS_and_conc_Au_NaCl} (a),
we examine the dependence of the differential capacitance on the $\ce{NaCl}$ electrolyte concentration.
As the concentration decreases, the differential capacitance systematically decreases as well.
Notably, the divergence observed at negative electrode potentials is strongly suppressed at lower concentrations and, in particular,
it becomes negligible at $0.1\,\text{M}$ and is no longer present at $0.05\,\text{M}$.

%
%
%
%
%
%
%
%
%
%
%
%
%
%
As we discussed in the previous section \ref{Ionic density profiles},
the pair-specific metal-cation parametrization plays an important role in determining the density profiles.
In Fig.~\ref{diff_cap_DRISM_PS_and_conc_Au_NaCl} (b),
we therefore analyze the dependence of the differential capacitance on the pair-specific $\ce{Au}$-$\ce{Na+}$ \ac{LJ} parameters,
again using the $\ce{Au}$ Heinz / mSPCE / $\ce{NaCl}$ Joung parametrization as a reference.

The black curve corresponds to the default \ac{LB} mixing rules.
For the other curves, we selectively reduce the $\varepsilon$ parameter by two orders of magnitude and progressively increase $\sigma$ by $0.2,\text{\AA}$ for the $\ce{Au}$-$\ce{Na+}$ pair.
Upon weakening the attractive interaction (smaller $\varepsilon$),
the differential capacitance at the \ac{PZC} decreases markedly.
A further increase of the repulsive radius $\sigma$ leads to a progressive reduction of the capacitance at negative potentials, whereas the positive branch remains largely unaffected.
This highlights the crucial role of the $\ce{Au}$-$\ce{Na+}$ interaction in determining the differential capacitance,
and the limitation of the \ac{LB} mixing rules.

\begin{figure}[!htb]
\begin{center}
\includegraphics[width=8.6cm]{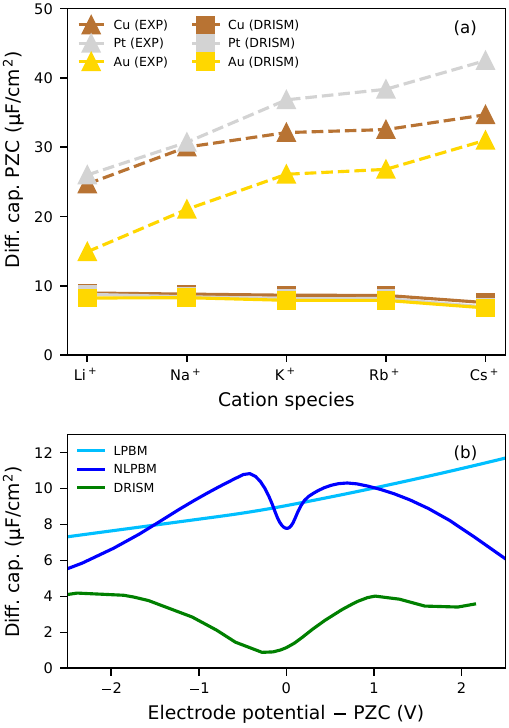}
\caption{
(a) Differential capacitance at the \ac{PZC} for the $(111)$ facet of $\ce{Cu}$, $\ce{Pt}$, and $\ce{Au}$ for different cationic species at $0.05\,\text{M}$ concentration
(adopting the Heinz / mSPCE / ion Joung parametrization)
The experimental data (triangle markers) \cite{garlyyevInfluenceNatureAlkali2018, xueHowNatureAlkali2020} show an increasing trend with increasing cationic species.
The computed differential capacitance from \ac{DRISM} (square markers), in contrast, is mostly constant, between 8 and $9\,\upmu \text{F}/\text{cm}^2$.
(b) Differential capacitance for the $\ce{Au}(111)$ facet in contact with $0.01\,\text{M}$ aqueous electrolyte.
In green, the differential capacitance from \ac{DRISM} for $\ce{Au}$ Heinz / mSPCE / $\ce{KBr}$ Joung parameters.
$\ce{KBr}$ leads to better convergence of the Fermi level (probably due to its larger ionic radius), so we present it instead of $\ce{NaCl}$.
The differential capacitance from the linear (light blue) and non-linear (blue) PB models is also shown for comparison.
}
\label{diff_cap_exp_VS_DRISM_and_0.01M}
\end{center}
\end{figure}

%
%
%
%
%
%
%
%
%
%
%
%
%
%
We next examine the influence of the cation species on the differential capacitance.
In Fig.~\ref{diff_cap_exp_VS_DRISM_and_0.01M}~(a),
we compare the differential capacitance at the \ac{PZC} predicted by DRISM for different electrodes in contact with various electrolytes at a concentration of $0.05,\text{M}$,
with the experimental measurements reported by \citet{garlyyevInfluenceNatureAlkali2018} and \citet{xueHowNatureAlkali2020}.
Experimentally, the differential capacitance at the \ac{PZC} increases along the series $\ce{Li+}$, $\ce{Na+}$, $\ce{K+}$, $\ce{Rb+}$, and $\ce{Cs+}$,
a trend attributed to the corresponding increase in cation hydration energy.
We note, however, that the experimental values for strongly adsorbing metals such as $\ce{Pt}$ should be interpreted with caution, 
as different measurement techniques can yield considerably different capacitances at the \ac{PZC} as recently discussed by \citet{frohlichProgressPitfallsMeasuring2025}.

In contrast to experiment, the DRISM results do not show a pronounced cation-dependent trend with the chosen parameterization:
the predicted differential capacitances lie within a narrow range of $8$-$9,\upmu\text{F/cm}^2$ for all cation species considered.
For completeness we also highlight, however,
that a recent experimental study by \citet{adnanTrackingSurfaceStructure2024a} on $\ce{Au}(111)$ similarly reports no significant dependence of the differential capacitance at the PZC on the identity of the cation,
suggesting that further experimental clarification is required before a final conclusion can be drawn.

%
%
%
%
%
%
%
%
%
%
%
%
%
%
%
To further assess the ability of \ac{DRISM} to capture the non-linear behavior of the differential capacitance near the \ac{PZC},
we also examined a lower electrolyte concentration of $0.01~\text{M}$.
At this concentration, the simulation requires a significantly larger cell,
which increases the fluctuations in the computed Fermi level; consequently, stronger smoothing is necessary.
Therefore, the resulting shape of the differential capacitance should be interpreted in a qualitative manner (see the SI).
In Fig.~\ref{diff_cap_exp_VS_DRISM_and_0.01M}~(b), we show the resulting comparison with the \ac{PB} model.

Despite the diffuse layer being in qualitative agreement with the \ac{PB} model (Fig.~\ref{Stern_and_diffuse_layer}), \ac{DRISM} predicts a lower capacitance than both linear \ac{PB} and non-linear \ac{PB}, which may be linked to a reduced Helmholtz layer contribution.
The characteristic non-linear behavior around the \ac{PZC} seems nevertheless to be captured at a qualitative level.
We also observe that the minimum in the differential capacitance is slightly shifted toward negative potentials rather than aligning with the \ac{PZC}.
However, this deviation may arise from the fitting procedure rather than reflecting a physical interfacial effect
(see the charging curve and the respective fit in the Fig.~S5).

%
%
%
%
%
%
%
%
%
%
%
%
%
%
%
\subsection{Adsorption Energies}
\label{Adsorption Energies}

Finally we analyze the impact of the DRISM parametrization on adsorption energies using $\ce{CO}$ as a prototypical probe, motivated by its central role in the electrocatalytic $\ce{CO2}$ reduction reaction (e$\ce{CO2}$RR).

\begin{figure}[htb]
\begin{center}
\includegraphics[width=8.6cm]{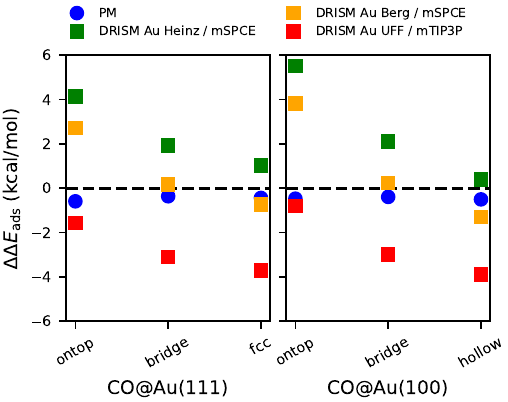}
\caption{
Solvation contribution to the adsorption energy of $\ce{CO}$ at the $\ce{Au}(111)$ and $\ce{Au}(100)$ facets, at the \ac{PZC}, for different \ac{DRISM} parametrizations and for the Poisson model in \vaspsolpp{}.
}
\label{dE_ads_sol}
\end{center}
\end{figure}

In Fig.~\ref{dE_ads_sol} we present the adsorption energies at the \ac{PZC} in solvent water only, for different parametrizations,
and we compare them to the one computed with the Poisson model in \vaspsolpp{}.
To highlight the effect of solvation, we take the adsorption energy in vacuum as an internal reference.
For the Poisson model we observe a minimal stabilization with respect to the vacuum reference,
and no significant trend with respect to the adsorption site.
These findings are in agreement with the previous study of \citet{heenenSolvationMetalWater2020}, where the solvation contribution to the $\ce{CO}$ adsorption energy was found to be small and relatively independent of the adsorption site.
On the other hand, for \ac{DRISM} we observe a significant dependence on the parametrization, with a difference of up to $\approx 6\,\text{kcal/mol}$.
The parametrization from Heinz is found to destabilize the $\ce{CO}$ adsorption, whereas the UFF parametrization has a stabilizing effect.
This can be explained by the $\ce{Au}$-$\ce{O}$ \ac{LJ} parametrization;
for the Heinz / mSPCE parametrization, the metal--oxygen interaction energy is $\varepsilon_{\ce{Au}-\ce{O}} \approx 0.9\,\text{kcal/mol}$,
while for the UFF / mTIP3P parametrization it is much lower $\varepsilon_{\ce{Au}-\ce{O}} \approx 0.08\,\text{kcal/mol}$.
The $\ce{CO}$ adsorption is disrupting the structure of the double layer at the interface, and the stronger $\ce{Au}$-$\ce{O}$ interaction in the Heinz parametrization leads to a larger solvation penalty for the adsorption.
This findings are consistent with the previous studies \cite{pasumarthiComparativeStudyElectrical2023,demeyereComparisonModernSolvation2024}.
A weaker adosorption energy for $\ce{CO}$ in electrolyte was also observed experimentally \cite{cuiDeterminingCOAdsorption2025} and ascribed to the competition between $\ce{CO}$ and water for the adsorption sites.
We also tested the parametrization from \citet{bergEvaluationOptimizationInterface2017} which is similar to the Heinz parametrization for the $\ce{Au}$-$\ce{O}$ interaction,
but introduces a predominantly repulsive $\ce{Au}$-$\ce{H}$ interaction, which decrease the solvation energy of $\approx 1\,\text{kcal/mol}$.
We further notice that the difference between the three \ac{DRISM} parametrizations is mostly constant and shows an analogous trend of increasing stabilization in the order: ontop, bridge, and fcc/hollow.

\begin{figure}[!htb]
\begin{center}
\includegraphics[width=8.6cm]{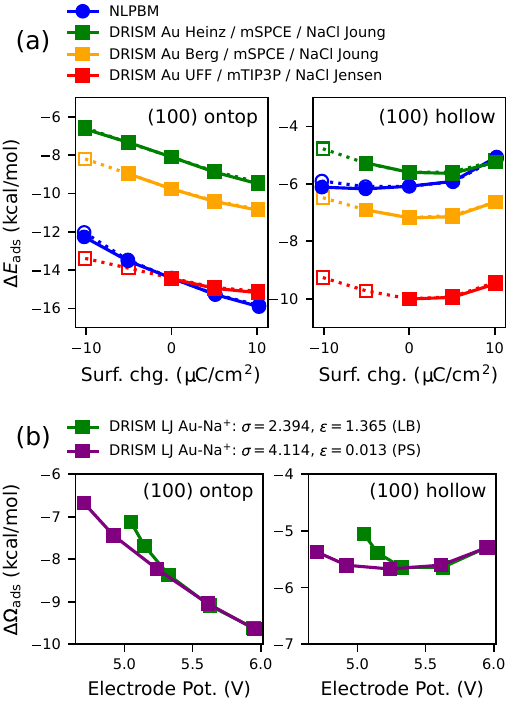}
\caption{
(a) Adsorption energy of $\ce{CO}$ at the $\ce{Au}(100)$ facet in contact with $1\,\text{M}$ electrolyte, as a function of the surface charge density.
For the hollow markers, the adsorption energy is computed from single point calculations, with structures optimized at the \ac{PZC}, since for some datapoints the full convergence of the atomic forces was not achieved, as discussed in the main text.
(b) Grand canonical adsorption energy of $\ce{CO}$ at the $\ce{Au}(100)$ facet in contact with $1\,\text{M}$ $\ce{NaCl}$ electrolyte,
as a function of the electrode potential (relative to the electrolyte bulk),
using the default \ac{LB} mixing rules (green),
and for a reduced $\ce{Au}$-$\ce{Na+}$ interaction (purple).
}
\label{E_adsorption_el}
\end{center}
\end{figure}

To test the effect of the mobile electrolyte ions, we computed the adsorption energies in $1\,\text{M}$ $\ce{NaCl}$ electrolyte, for two different surface sites $\ce{CO}@\ce{Au}(100)$ ontop and hollow.
In Fig.~\ref{E_adsorption_el}~(a), we show the dependence of the adsorption energy on the surface charge density.

We highlight that for some datapoints, it was not possible to achieve full convergence of the atomic forces under the defined threshold value.
We ascribe this behavior to the strong accumulation of $\ce{Na+}$ close to the interface at negative surface charge densities, which leads to a very steep energy landscape and makes the optimization process more difficult.
Therefore, we also report the adsorption energy computed from single point calculations, with structures optimized at the \ac{PZC}, with hollow markers and dotted lines.

Given the small differences between the two approaches, we regard the overall trends as reliable, even for the points where full convergence was not achieved.
The different parametrizations show a very similar behavior, with analogous solvation trends as observed in Fig.~\ref{dE_ads_sol}.
Similar trends are also observed for the \ac{PB} model.

A different picture emerges when we examine the dependence of the adsorption energy on the electrode potential, as shown in Fig.~\ref{E_adsorption_el}~(b).
In green we show the Heinz / mSPCE / $\ce{NaCl}$ Joung parametrization, with the conventional \ac{LB} mixing rules,
while in purple we show the same parametrization but with a reduced $\ce{Au}$-$\ce{Na+}$ interaction, as in Fig.~\ref{diff_cap_DRISM_PS_and_conc_Au_NaCl}.
The asymmetry encountered for the Heinz /mSPCE / $\ce{NaCl}$ Joung parametrization, in the capacitance curve at negative potentials is now reflected in the adsorption energy.
When decreasing the $\ce{Au}$-$\ce{Na+}$ interaction, the adsorption energy becomes more symmetric with respect to the \ac{PZC},
and we regard it as more accurate,
since, as discussed in section~\ref{differential capacitance}, the steep increase of the differential capacitance at negative potentials is not observed experimentally.

%
%
%
%
%
%
%
%
%
%
%
%
%
%
%
%
%
%
%
%
\section{Conclusion}
\label{conclusion}

In this work, we assessed the dielectrically consistent \ac{DRISM} embedding for modeling the \ac{EDL} and compared its predictions with explicit \ac{MD} data and continuum \ac{PB} models.
For the gold--water interface, \ac{DRISM} reproduces the main features of the solvent density profiles and yields a qualitatively reasonable microscopic description of the interfacial liquid structure.
For ionic density profiles, the method also captures the position of the main peaks observed in explicit \ac{MD} simulations.
Additionally, our results show that the predicted structure of the compact layer depends strongly on the parametrization of the metal--ion interactions.
With the default \ac{LB} mixing rules, $\ce{Na+}$ can accumulate too strongly in the \ac{IHP}, which leads to divergent differential capacitance at negative electrode potentials for concentrated $\ce{NaCl}$ electrolytes, a behavior that we regard as unphysical.
We show that by introducing pair-specific $\ce{Au}$-$\ce{Na+}$ parameters, this excessive cation accumulation is reduced, and this is reflected in a more symmetric differential capacitance curve.

In the dilute limit, \ac{DRISM} recovers the expected exponential decay of the diffuse charge distribution, and charge profiles lie between the linear and non-linear \ac{PB} model.

For $\ce{CO}$ adsorption, the solvation contribution depends strongly on the chosen metal--water parameters, with differences up to $6\,\text{kcal/mol}$ but the trends across different adsorption sites are consistent across different parametrizations.
The dependence of the adsorption energy on the electrolyte is more evident when examining the adsorption energy as a function of the electrode potential, where the asymmetry in the adsorption energy follows the same behavior observed in the differential capacitance.
Also in this case we suggest that the introduction of pair-specific parameters for the $\ce{Au}$-$\ce{Na+}$ interaction leads to a more symmetric behavior of the adsorption energy with respect to the \ac{PZC}, which we regard as more accurate.

Overall, our results demonstrate that the adoption of pair-specific parameters within the \qmdrism framework can improve the flexibility of the model.
We believe that a systematic optimization of the pair-wise parameters, could significantly enhance the accuracy of the model and can be objective of future work.

%
%
%
%
%
%
%
%
%
%
%
%
%
%
%
%
%
%
\section{Methods}
\label{methods}
%
%
%
%
%
%
%
%
%
%
%
%
%
%
%
%
We used the Vienna \textit{ab initio} Simulation Package \cite{kresseEfficientIterativeSchemes1996} (VASP) with the \vaspsolpp \cite{islamImplicitElectrolyteModel2023} Poisson--Boltzmann (PB) model and Quantum ESPRESSO \cite{giannozziQUANTUMESPRESSOModular2009, giannozziAdvancedCapabilitiesMaterials2017a} with the built-in implementation of the reference interaction-site model (RISM) \cite{nishiharaHybridSolvationModels2017, hagiwaraDevelopmentDielectricallyConsistent2022}.
For the calculations with VASP, we builtin PAW pseudopotentials while with Quantum ESPRESSO, we used the PAW pseudopotentials from the PSlibrary \cite{dalcorsoPseudopotentialsPeriodicTable2014}.
The initial atomic structures were generated using the Atomic Simulation Environment (ASE) \cite{hjorthlarsenAtomicSimulationEnvironment2017}.

%
%
%
%
%
%
%
%
%
%
%
%
%
%
\textbf{Density Profiles and Differential Capacitance}
For computing the density profiles and the differential capacitance, we chosen a $1 \times 1 \times 6$ slab model, with the default experimental lattice parameters in ASE.
To ensure that the cell size would not affect the results, we compared the density profiles obtained from a slab with a larger surface $3 \times 3$.
As shown in Fig.~S1, the profiles overlap well at both high and low electrolyte concentrations, proving that a $1 \times 1$ surface cell is sufficient to investigate the \ac{EDL}'s properties within the \qmesmrism{} framework.
We employed the PBE density functional \cite{perdewGeneralizedGradientApproximation1996} and Gaussian smearing with a broadening parameter of $0.2\,\text{eV}$ ($\equiv 0.0147\,\text{Ry}$).
We employed a plane wave cutoff of $500\,\text{eV}$ ($\equiv 36\,\text{Ry}$, with an additional $\times 8$ cutoff for the electron density in Quantum ESPRESSO).
For sampling the Brillouin zone, we used a $11 \times 11 \times 1$ $\Gamma$-centered k-point grid.
For the VASP calculations, we used an energy threshold of $10^{-6}\,\text{eV}$ for the electronic energy, while for Quantum ESPRESSO we set the threshold to $10^{-8}\,\text{Ry}$.
For RISM, two additional thresholds must be defined:
the first establishes the convergence of the 1D-RISM calculation, for which we used a value of $10^{-8}\,\text{Ry}$;
the second is related to the 3D-RISM calculation, which we set to $10^{-7}\,\text{Ry}$.
We used a tighter convergence than previously reported \cite{haruyamaElectrodePotentialDensity2018}, since we found the Fermi level to be very sensitive to the 3D-RISM convergence.
We report, however, that for the mTIP5P water model, it was not possible to reach the desired convergence, and therefore, we increased the threshold to $10^{-6}\,\text{Ry}$.
Finally, for RISM, an additional plane wave cutoff is required to expand the 3D particle densities; we used $300\,\text{Ry}$.

In both the implicit electrolyte models, the electrode potential is defined with respect to the electric potential in the electrolyte bulk, according to the following expression:
\begin{equation}
\Phi_{\text{el.de}}
=
\phi_{\text{el.yte}}
-
\frac{\bar{\mu}_{\mathrm{e}}}{q_{\mathrm{e}}}
\end{equation}
Where $\bar{\mu}_{\mathrm{e}}$ is the electrochemical potential of the electrons in the metal electrode, i.e., the Fermi level, $q_{\mathrm{e}}$ is the elementary charge, and $\phi$ is the electric potential in the electrolyte bulk region.

The differential capacitance is computed by performing several calculations at different surface charges, i.e., by changing the number of electrons in the system.
The surface charge is defined as the excess of electronic charge in the system divided by the surface area of the cell.
For the \ac{PB} model implementation in VASP, since the electrolyte is present on both sides of the cell, the surface area must be multiplied by two.
The \esmrism{} implementation in Quantum ESPRESSO allows for a non-periodic vacuum/slab/electrolyte setup; therefore, the surface area is just the one in contact with the electrolyte.
The resulting charging curve is fitted with a cubic spline and differentiated to obtain the differential capacitance.
A different level of smoothing might be required, given that even small fluctuations in the computed Fermi level lead to large discontinuities in the first derivative (as previously reported \cite{demeyereComparisonModernSolvation2024}). 

%
%
%
%
%
%
%
%
%
%
%
%
%
%
\textbf{Adsorption Energies}
For the adsorption energy we used the RPBE density functional \cite{hammerImprovedAdsorptionEnergetics1999} along with the D3 dispersion correction (with ``zero damping'' function) \cite{grimmeConsistentAccurateInitio2010} following previous studies \cite{heenenSolvationMetalWater2020,dudzinskiFirstStepOxygen2023a} on the gold electrode interface.
We used the plane wave cutoff of $600\,\text{eV}$ ($\equiv 44\,\text{Ry}$, with an additional $\times 8$ cutoff for the electron density in Quantum ESPRESSO).
To construct the slab we employed an optimized lattice constant for a $1 \times 1 \times 1$ cubic cell (with a $21 \times 21 \times 21$ k-point mesh).
The resulting lattice constants are $4.172\,\text{\AA}$ for VASP, and $4.196\,\text{\AA}$ for Quantum ESPRESSO.
The small difference can be ascribed to the different pseudopotentials.
The slab model consists of a $3 \times 3$ surface cell with four layers, where the two bottom layers are fixed to the bulk positions and the two top layers are allowed to relax.
The Brillouin zone is sampled with a $3 \times 3 \times 1$ $\Gamma$-centered k-point grid, and Gaussian smearing with a broadening parameter of $0.1\,\text{eV}$ ($\equiv 0.0073\,\text{Ry}$) is used.
The force convergence threshold for the geometric optimization is set to $0.01\,\text{eV/\AA}$ for VASP and $0.002\,\text{Ry/Bohr}$ for Quantum ESPRESSO. As we highlight on the main text, for some of the Quantum ESPRESSO calculations, denoted with the hollow markers in Fig.~\ref{E_adsorption_el}~(a), was not possible to achieve the desired convergence of the forces, and therefore, we report the adsorption energy computed from single point calculations, with the geometry optimized at the \ac{PZC}.
The adsorption energy is computed as:
\begin{equation}
\Delta E_{\text{ads}} = E_{\ce{CO}@\text{slab}}
-
E_{\text{slab}}
-
E_{\ce{CO}}
\end{equation}
The solvation contribution to the adsorption energy is then computed as the difference between the adsorption energy in solvent and the adsorption energy in vacuum.
\begin{equation}
\Delta \Delta E_{\text{ads}} = \Delta E_{\text{ads,solvent}}
-
\Delta E_{\text{ads,vacuum}}
\end{equation}
The for the Grand-canonical adsorption energy, we also include the contribution from the change in the number of electrons in the system, as follows:
\begin{equation}
\Delta \Omega_{\text{ads}}
=
\Delta E_{\text{ads}}
-
\Delta N_{\mathrm{e}} \cdot \bar{\mu}_{\mathrm{e}}
\end{equation}
We point out however, that for computational convenience, we did not include the vibrational contribution to the adsorption energy.
Since our main focus is on the solvation contribution, we expect this approximation to be reasonable,
and not affect the overall trends and discussion.

%
%
%
%
%
%
%
%
%
%
%
%
\textbf{Electrolyte Models}
The main parameters of the \ac{PB} model are the dielectric constant of the solvent, in our case water,
the radius of the dielectric cavity, the radius of the ionic cavity, and the maximum ionic packing diameter.
We used the default settings, as suggested in the original implementation \cite{islamImplicitElectrolyteModel2023}.

Within the \ac{RISM} formalism, the pairwise potentials among the interaction sites must be provided.
In the specific implementation by \citet{nishiharaHybridSolvationModels2017}, the pairwise potentials are given in the form of Lennard-Jones 12--6 expression and the Lorentz--Berthelot combination rules are adopted $\sigma_{AB} = \frac{\sigma_{AA} + \sigma_{BB}}{2}$, $\epsilon_{AB} = \sqrt{\epsilon_{AA} \cdot \epsilon_{BB}}$.
The use of pair-specific parameters for the \ac{LJ} interaction between the explicit system and the implicit electrolyte is described case by case in the main text (to define pair-specific parameters we used our own modified version of Quantum ESPRESSO, since at the present moment only the Lorentz--Berthelot mixing rules are implemented in the official release).

For water, we use the SPCE \cite{berendsenMissingTermEffective1987}, the TIP3P \cite{jorgensenComparisonSimplePotential1983}, and the TIP5P \cite{mahoneyFivesiteModelLiquid2000} models, with the modifications adopted in \cite{haruyamaElectrodePotentialDensity2018}.
The modifications consist of additional \ac{LJ} potentials for hydrogen atoms and lone charge pairs in the TIP5P model.
The repulsive part of the \ac{LJ} potential is needed to compute site-site particle correlations, which would otherwise diverge at short distances.
To highlight this difference from the original models, we denote them as ``modified'': mSPCE, mTIP3P, and mTIP5P.
For the ions, we used the parameters optimized for the SPCE water model from the work of \citet{joungDeterminationAlkaliHalide2008}, and the OPLS parametrization from \citet{jensenHalideAmmoniumAlkali2006} in combination with the TIP3P and TIP5P water models.
For the explicit \ac{QM} system, we used the \ac{LJ} parameters from the \ac{UFF} \cite{rappeUFFFullPeriodic1992} and from \citet{heinzAccurateSimulationSurfaces2008}.

Finally, for the dielectric correction in \ac{DRISM}, we used the same settings as in the original implementation \cite{hagiwaraDevelopmentDielectricallyConsistent2022}.

%
%
%
%
%
%
%
%
%
%
%
%
%
%
%
%
%
%
%
%
%
%





\begin{suppinfo}

Additional details and figures on the density profiles, the differential capacitance, and the adsorption energies are provided in the Supporting Information.

\end{suppinfo}

\bibliography{bibliography}

\end{document}